\documentclass[aps,pre,twocolumn,showpacs,superscriptaddress,groupedaddress]{revtex4}

\usepackage{latexsym}
\usepackage[dvips]{graphics}
\usepackage{amssymb,amstext,amsfonts,amsmath}

\usepackage[T1]{fontenc} 
\usepackage[utf8x]{inputenc}
\usepackage[usenames,dvipsnames]{color}

\usepackage{natbib} 
\date{\today}

\begin{document}


\title{Time-dependent solutions for a stochastic model of gene expression with molecule production in the form of a compound Poisson process}




\author{Jakub Jędrak}
\email[]{jjedrak@ichf.edu.pl}
\affiliation{Institute of Physical Chemistry, Polish Academy of Sciences, ul. Kasprzaka 44/52, 01-224 Warsaw, Poland}
\author{Anna Ochab-Marcinek}
\affiliation{Institute of Physical Chemistry, Polish Academy of Sciences, ul. Kasprzaka 44/52, 01-224 Warsaw, Poland}

\date{\today}

\begin{abstract}
\vspace{0.3cm}
We study a stochastic model of gene expression, in which protein production has a form of random bursts whose size distribution is arbitrary, whereas protein decay is a first-order reaction. We find exact analytical expressions for the time evolution of the cumulant-generating function for the most general case when both the burst size probability distribution and the model parameters depend on time in an arbitrary (e.g. oscillatory) manner, and for arbitrary initial conditions. We show that in the case of periodic external activation and constant protein degradation rate, the response of the gene is analogous to the RC low-pass filter, where slow oscillations of the external driving have a greater effect on gene expression than the fast ones. We also demonstrate that the $n$-th cumulant of the protein number distribution depends on the $n$-th moment of the burst size distribution. We use these results to show that  different measures of noise (coefficient of variation, Fano factor, fractional change of variance) may vary in time in a different manner. Therefore, any biological hypothesis of evolutionary optimization based on the nonmonotonicity of a chosen measure of noise must justify why it assumes that biological evolution  quantifies noise in that particular way. Finally, we show that not only for exponentially distributed burst sizes but also for a wider class of burst size distributions (e.g. Dirac delta and gamma) the control of gene expression level by burst frequency modulation gives rise to proportional scaling of variance of the protein number distribution to its mean, whereas the control by amplitude modulation implies proportionality of protein number variance to the mean squared.

\end{abstract}
\maketitle

\section{Introduction \label{Intro}}

It has been confirmed experimentally that in living cells both mRNA \cite{golding2005real} and protein \cite{ozbudak2002regulation, cai2006stochastic, yu2006probing, taniguchi2010quantifying, choi2008stochastic} production may take form of stochastic bursts of a random size. The presence of bursts may be a result of processes involving short-lived molecules (e.g. the mRNA in case of protein production), concentration of which may be treated as a fast degree of freedom \cite{friedman2006linking,shahrezaei2008analytical}. The number of protein molecules that can be produced from a single mRNA molecule before the latter is degraded is a random variable, and its distribution may, in the several experimentally known cases, be well approximated by geometric or exponential distribution \cite{cai2006stochastic, yu2006probing, taniguchi2010quantifying}. For that reason, in most of the existing models of bursty gene expression, the exponential (or geometric in a discrete case) bursts of protein \cite{paulsson2000random, friedman2006linking, shahrezaei2008analytical, aquino2012stochastic, lin2016bursting} or mRNA \cite{aquino2012stochastic} production are considered.

However, in the case of eukaryotic cells, certain models predict nonexponential distributions of burst sizes \cite{schwabe2012transcription,elgart2011connecting,kuwahara2015beyond}. In particular, in the case of transcriptional bursts the molecular ratchet model predicts peaked distributions, that resemble gamma distribution \cite{schwabe2012transcription}. Therefore, it seems desirable to study the analytically tractable models of bursty gene expression dynamics with a general, nonexponential form of burst size distributions. 

Also, for the majority of stochastic models of gene expression proposed to date, even if the time-dependent solutions are considered \cite{iyer2009stochasticity, tabaka2010binary, ramos2011exact, feng2012analytical, pendar2013exact, kumar2014exact}, it is usually assumed that model parameters are time-independent. However, taking into account the time variation of the model parameters, in particular the periodic time dependence of the rate of protein production \cite{mugler2010information} gives us an opportunity to model in a simple manner the response of a genetic circuit to oscillatory regulation and to indicate some qualitative properties of solutions for other oscillating parameters.

In this paper, we investigate a simple gene expression model, which is a natural generalization of the analytical framework proposed in Ref. \cite{friedman2006linking}, and which may serve as a model of both transcription and translation \cite{aquino2012stochastic}. Namely, in contrast to Ref. \cite{friedman2006linking} we consider the case of an arbitrary (not necessarily exponential) burst size probability distribution and time-dependent model parameters. However, gene autoregulation is neglected. To the best of our knowledge, the time-dependent solutions of the model of Ref. \cite{friedman2006linking} have not been known to date even in the absence of gene autoregulation or for the simplest case of time-independent model parameters. 

We find the explicit time dependence of the cumulant-generating function for the probability distribution of molecule (protein) concentration. This general result is then applied to describe the oscillatory response of a gene to periodic modulation of the rate of protein production. In particular, we consider a gene driven by a single-frequency, sinusoidal regulation. In such a case, the time dependence of the mean molecule concentration consists of both the transient, exponentially decaying part and of the periodic part, whose amplitude depends on the driving frequency.
We also {point out} that the division of the system's response into periodic and transient part remains true in a more general case, when the model parameters are periodic functions of time.

We also show a simple relationship that links the $n$-th cumulant of the protein number distribution and the $n$-th moment of the burst size distribution. In particular, this relationship is proportional in the steady state. We use these results to discuss the question of possible evolutionary optimization of cellular processes with respect to noise intensity. Since it has been shown experimentally that distributions of protein numbers have universal scaling properties (variance proportional to mean or variance proportional to mean squared) \cite{salman2012universal,bar2006noise}, we use our results to gain an insight into possible origins of such scalings in the properties of the burst size distributions.  
 
Stochastic models with bursty dynamics similar to the model considered here are known both in mathematics (so-called Takacs processes \cite{cox1986virtual,takacs1961transient}) and in physics (under the name of compound Poisson processes), where such models are used not only to describe stochastic dynamics of transcription or translation, but also to model such diverse phenomena as diffusion with jumps \cite{luczka1995white,czernik1997thermal,luczka1997symmetric}, time dependence of soil moisture \cite{porporato2004phase,daly2006probabilistic,daly2010effect,suweis2011prescription,mau2014multiplicative}, dynamics of snow avalanches \cite{perona2012stochastic}, statistics of the solar flares \cite{wheatland2008energetics,wheatland2009monte} and oil 
prices on the stock market \cite{askari2008oil}. And therefore, our results may be relevant to other fields beyond stochastic modeling of gene expression.

\section{Results \label{Results}}
Let us consider a source (gene) that creates objects (protein or mRNA molecules) of a single type, denoted by X, which are subsequently degraded or diluted due to the system size expansion, e.g. cell growth and division, 
\begin{eqnarray}
\text{DNA} & \xrightarrow{I(t)} &   \text{X}, ~~~~~
\text{X}  \xrightarrow{\gamma(t)} \emptyset. 
\label{biochemical reaction scheme}
\end{eqnarray}
We focus on the simplest situation, when the molecules interact neither with each other, nor with the source. In consequence, the probability of degradation of a single molecule does not depend on the total number of molecules in the system. This assumption leads to a linear decay process (first-order reaction), which is the simplest, but arguably the most natural choice here. Still, we assume that both the source intensity ${I(t)}$ and the decay parameter $\gamma(t)$ may vary with time in an arbitrary manner. Therefore, although we assume that the characteristics of the source are independent on the number of molecules present in the system (feedback effects are neglected), we allow the source (gene) to be externally regulated. If the number of molecules is sufficiently large, the continuous approximation is justified  and the molecule concentration may be used instead of the exact copy number of molecules.

In order to obtain the stochastic description of the system, we assume that the molecule production takes the form of bursts of random size. Namely, the number of newly created molecules (or the magnitude of a concentration jump in the present continuous model), $u$, is a stochastic variable drawn from the probability distribution $\nu(u,t)$, which may be explicitly time-dependent. It is assumed here, that burst duration is short enough that even large bursts can be treated as instantaneous. The time of appearance of each burst is also a random variable.

The occurrence of stochastic bursts in a given system may be due to the presence of some processes that are much faster than production or degradation of molecules in question; such processes are not explicitly taken into account within the model. For example, translational bursts of proteins are attributed to the existence of short-lived mRNA molecules \cite{shahrezaei2008analytical, lin2016bursting}. However, it is not our aim here to relate the functional form of the burst size probability distribution to dynamics of fast degrees of freedom. Rather, we treat bursty dynamics as a well-justified approximation leading to reasonable effective description of the system at hand. 

The deterministic model describing the kinetics of reactions (\ref{biochemical reaction scheme}) is given by a simple rate equation (\ref{general deterministic ode}), see Appendix \ref{Appendix B}. Its stochastic counterpart is the following Langevin-like equation 
\begin{eqnarray}
\dot{x} &=& I(t) - \gamma(t) x, 
\label{langevin ode}
\end{eqnarray}
where $x \geq 0$ is the molecule concentration and dot denotes the time derivative. $I(t)$ appearing in (\ref{langevin ode}) is {now} a compound Poisson process, i.e.
\begin{eqnarray}
I(t) &=& \sum_{k=1}^{N(t)} u_k \delta(t-t_k), 
\label{compound poisson process}
\end{eqnarray}
where $u_k$ is the size of the molecule burst (concentration jump) that takes place at $t=t_k$ and $N(t)$ is the number of concentration jumps in the interval $[0, t)$. 

Stochastic differential equations similar to (\ref{langevin ode}) have been used to model diffusion in asymmetric periodic potentials \cite{luczka1995white,czernik1997thermal,luczka1997symmetric}, soil moisture dynamics and other phenomena in geophysics \cite{perona2012stochastic, porporato2004phase, daly2006probabilistic,daly2010effect,suweis2011prescription,mau2014multiplicative}, astrophysics \cite{wheatland2008energetics,wheatland2009monte} and economics \cite{askari2008oil}.


Instead of Eq. (\ref{langevin ode}) it is more convenient to study the corresponding master equation\footnote{According to the terminology of Ref. \cite{gardiner2009stochastic}, Eq. (\ref{unregulated t dependent ME of Friedman}) is a special case of differential Chapman-Kolmogorov equation.}, proposed in Ref. \cite{friedman2006linking}  
\begin{eqnarray}
\frac{\partial p(x,t)}{\partial t} &=& \gamma(t)\frac{\partial }{\partial x}\left[ x p(x,t)\right] \nonumber \\ &+& k(t) \int_{0}^{x} w(x-x^{\prime}, t)p(x^{\prime},t) dx^{\prime}.
\label{unregulated t dependent ME of Friedman}
\end{eqnarray}
In the above equation, $p(x,t)$ is a time-dependent probability distribution of molecule concentration in the population of cells. We also have 
\begin{equation}
w(u, t) = \nu(u, t) - \delta(u),
\label{definition of w(u)}
\end{equation}
where $\nu(u,t)$ is the burst size probability distribution, $\delta(u)$ denotes Dirac delta distribution, $u=x-x^{\prime}$ is the burst size, whereas $\gamma(t)$ and $k(t)$ are time-dependent model parameters (in Ref. \cite{friedman2006linking}, only time independent model parameters have been considered). 

Note that from Eq. (\ref{unregulated t dependent ME of Friedman}) one can obtain equations for the time evolution of moments of $p(x, t)$, see Appendix \ref{Appendix C}. However, solution of the moment equations is tedious, and it is usually much more convenient to work with the moment generating function.

In order to solve Eq. (\ref{unregulated t dependent ME of Friedman}), we apply the Laplace transform: $p(x,t)\to \hat{p}(s,t)=\mathcal{L}\{p(x,t)\}$, $w(u,t)\to \hat{w}(s,t)=\mathcal{L}\{w(u,t)\}$, i.e., $\hat{w}(s,t)=\hat{\nu}(s,t)-1$. In result, Eq. (\ref{unregulated t dependent ME of Friedman}) is transformed into the following first-order linear partial differential equation 
\begin{equation}
\frac{\partial \hat{p}(s,t)}{\partial t} + \gamma(t) s\frac{\partial \hat{p}(s,t)}{\partial s} - k(t)\hat{w}(s,t)  \hat{p}(s,t) = 0,
\label{unregulated t dependent ME of Friedman in s space}
\end{equation}
which can be solved by the standard method of characteristics \cite{van2007stochastic, arfken2011mathematical}. We obtain
\begin{equation}
\hat{p}(s,t)=\Phi(\Omega(t)s) e^{\mathcal{G}\left(\Omega(t)s, t \right)},
\label{solution of the most general equation of characteristics}
\end{equation}
where 
\begin{equation}
\Phi(z)=\hat{p}(z, t_0)=\mathcal{L}\{p(x, t_0)\}
\label{definition of Phi}
\end{equation}
is the Laplace transform of the initial probability distribution $p(x, t_0)$;  
\begin{equation}
\Omega(t) = \exp \left(-\int_{t_0}^{t} \gamma(t^{\prime}) dt^{\prime}\right),
\label{definition of Omega and Gamma}
\end{equation}
whereas $\mathcal{G}(z,t)$ is defined as
\begin{equation}
\mathcal{G}(z,t)=\int_{t_0}^{t} k(t^{\prime})\hat{w}\left(\frac{z}{\Omega(t^{\prime})}, t^{\prime} \right)dt^{\prime}.
\label{definition of F}
\end{equation}
It can be easily verified that for $\hat{p}(s, t)$ (\ref{solution of the most general equation of characteristics}) we have  
\begin{equation}
\hat{p}(s, t_0)=\Phi(s), ~~~ \hat{p}(0, t)=1.
\label{normalization and initial condition general}
\end{equation}

If $k(t)$, $\gamma(t)$, and $\hat{w}(s,t)$ are periodic functions of time (including constant function treated as a special case of periodic function), and at least one of these three functions is not a constant function, time evolution of $p(x,t)$ has an oscillatory character. More precisely, it is shown that each cumulant of $p(x,t)$ consists of both the periodic part and the exponentially decaying transient terms, cf.  Appendix \ref{Appendix A}.

In most cases, $\hat{p}(s,t)$ given by (\ref{solution of the most general equation of characteristics}) cannot be expressed in terms of elementary or standard special functions. Even if for some choice of $k(t)$, $\gamma(t)$, and 
$\hat{w}(s,t)$ functions it is feasible to obtain a closed analytical formula for $\hat{p}(s,t)$, the analytical evaluation of the inverse Laplace transform and hence the explicit analytical form of $p(x,t)$ is usually out of question (a notable exception, for which the explicit form of $p(x,t)$ can be obtained is analyzed in Section \ref{Example exponential}). 

However, making use of the relationship between $\hat{p}(s,t)$, moment generating function $M(s, t)$ and the cumulant generating function $K(s, t)$,
\begin{equation}
\hat{p}(s,t)= M(-s, t) = \sum_{m=0}^{\infty} \mu_m(t) \frac{(-s)^m}{m!},
\label{pdf series expansion of Laplace transform gives moments}
\end{equation}
\begin{equation}
\ln[\hat{p}(s,t)] = K(-s, t) = \sum_{m=1}^{\infty} \kappa_m(t) \frac{(-s)^m}{m!},
\label{ln pdf series expansion of Laplace transform gives cumulants}
\end{equation}
one may find the exact analytical form of the time evolution of moments $\mu_r(t)$ and cumulants $\kappa_r(t)$  of $p(x,t)$ \cite{van2007stochastic, gardiner2009stochastic}.
The cumulants of $p(x, t)$ are of special interest here; from (\ref{solution of the most general equation of characteristics}), (\ref{definition of Phi}), (\ref{definition of Omega and Gamma}), (\ref{definition of F}), and (\ref{ln pdf series expansion of Laplace transform gives cumulants}) one gets
\begin{eqnarray}
\kappa_r(t) &=& (-1)^r \left(\frac{\partial^r \ln[\hat{p}(s,t)]}{\partial s^r}\right)_{s=0} \nonumber \\ &=& \left[\Omega(t)\right]^{r} \left( \kappa_r(0) + \int_{t_0}^{t} \frac{k(t^{\prime}) m_r(t^{\prime}) }{\left[\Omega(t^{\prime})\right]^{r}}  dt^{\prime}\right).
\label{cumulant most general}
\end{eqnarray}
{In the above equation, $m_r$ denotes $r$-th moment of the burst size probability distribution $\nu(u, t)$ (\ref{definition of w(u)}), i.e.,}
\begin{equation}
m_r(t) = \int_{0}^{\infty} u^r \nu(u, t) du.
\label{moments of ni def}
\end{equation}
From (\ref{cumulant most general}) we see that the time evolution of $\kappa_r(t)$ depends only on its initial value, $\kappa_r(0)$, on the time dependence of the model parameters $k(t)$, $\gamma(t)$, and on the time evolution of $r$-th moment of $\nu(u,t)$, but it does not depend {explicitly} on any other cumulants of $p(x,t)$ or moments of $\nu(u,t)$. Note that by using Eqs. (\ref{ln pdf series expansion of Laplace transform gives cumulants}) and (\ref{cumulant most general}) we can reconstruct (at least in principle) the time evolution of $\hat{p}(s,t)$, provided that the time evolution of all moments $m_r(t)$ of $\nu(u, t)$ as well as the initial distribution $p(x,0)$ are given.

Eq. (\ref{cumulant most general}) can also be obtained in an alternative way, which does not require the solution of Eq. (\ref{unregulated t dependent ME of Friedman in s space}). Namely, dividing  Eq. (\ref{unregulated t dependent ME of Friedman in s space}) by $\hat{p}(s,t)$ we obtain the following equation for $K(-s, t)=\ln[\hat{p}(s,t)]$ given by Eq. (\ref{ln pdf series expansion of Laplace transform gives cumulants})
\begin{equation}
\frac{\partial K(-s,t)}{\partial t} + \gamma(t) s\frac{\partial K(-s,t)}{\partial s} - k(t)\hat{w}(s,t)  = 0.
\label{unregulated t dependent ME of Friedman in s space for K}
\end{equation}
If we compute the $r$-th derivative of Eq. (\ref{unregulated t dependent ME of Friedman in s space for K}) with respect to $s$-variable, and subsequently put $s=0$, we get the time-evolution equation for $\kappa_r$ 
\begin{equation}
\dot{\kappa}_r(t) + r \gamma(t) \kappa_r(t) - k(t)m_r(t) = 0,
\label{time evolution of r-th cumulant derived from K}
\end{equation}
from which we immediately obtain (\ref{cumulant most general}).

The two most important cumulants are the mean molecule concentration $\kappa_1(t)=\mu_1(t)$ and variance $\kappa_2(t)$. In particular, $\kappa_1(t)$ is given by 
\begin{equation}
\kappa_1(t)=\Omega(t)\left[\kappa_1(0) + \int_{t_0}^{t} \frac{k(t^{\prime})m_1(t^{\prime})}{\Omega(t^{\prime})}dt^{\prime} \right],
\label{solution of time evolution equation for the first moment of p}
\end{equation}
cf. Eq. (\ref{solution of time evolution equation for the first moment of p appendix}) in Appendix \ref{Appendix C}. $\kappa_1(t)$ and $\kappa_2(t)$ are of special interest also with the connection with two standard noise measures frequently used in biology: the Fano factor $F$ and the coefficient of variation $\eta$, defined as
\begin{equation}
F(t)= \frac{\kappa_2(t)}{\kappa_1(t)}, ~~~~~\eta(t) = \frac{\sqrt{\kappa_2(t)}}{\kappa_1(t)}.
\label{Fano factor and eta}
\end{equation}
\subsection{Periodic gene regulation \label{oscillatory time dependence}}
Let us now analyze the case of a time-independent, but otherwise arbitrary burst size probability distribution $\nu_{}(u)$, constant decay rate $\gamma$ and molecule production rate (burst frequency) $k(t)$ of the form 
\begin{equation}
k(t) = C_1 \sin\left(\omega_f t + \varphi\right) + C_2,
\label{sinusoidal k}
\end{equation}
where $0\leq C_1 < C_2$. In other words, our gene is periodically driven with a single angular frequency, 
\begin{equation}
\omega_f = 2\pi /T,
\label{definition of omega f}
\end{equation}
where $T$ is an oscillation period; $\varphi$ is the initial phase. Making use of (\ref{cumulant most general}) and (\ref{sinusoidal k}), one can easily compute time evolution of $r$-th cumulant of $p(x, t)$. Assuming for simplicity $t_0=0$, we get  
\begin{eqnarray}
\kappa_r(t) &=& \kappa_r(0)e^{-r\gamma t} + \frac{C_2  m_r}{r \gamma}\left(1-e^{-r\gamma t}\right) \nonumber \\ &+& \frac{C_1 m_r \sin(\omega_f t + \varphi + \beta)}{\sqrt{r^2 \gamma^2 + \omega^2_f}} \nonumber \\ &-& \frac{C_1  m_r  \sin(\varphi+\beta)e^{-r\gamma t}}{\sqrt{r^2 \gamma^2 + \omega^2_f}} ,
\label{cumulant constant and periodic model parameters example sqrt}
\end{eqnarray}
%
where 
\begin{equation}
\beta = \arctan \left(\frac{-\omega_f}{r \gamma}\right).
\label{definition of beta angle}
\end{equation}
$\kappa_r(t)$ given by (\ref{cumulant constant and periodic model parameters example sqrt}) contains both the transient, exponentially decaying terms and the terms which are periodic functions of time, oscillating with an angular frequency of the driving. What is important, and easily visible when $\kappa_r(t)$ is written in a form (\ref{cumulant constant and periodic model parameters example sqrt}), the oscillation amplitude depends on both $\omega_f$ and $\gamma$,
\begin{equation}
A_r(\gamma, \omega_f)=\frac{C_1}{\sqrt{r^2 \gamma^2 + \omega^2_f}}. 
\label{amplitude sqrt}
\end{equation}
$A_r(\gamma, \omega_f)$ (\ref{amplitude sqrt}) is a monotonically decreasing function of $\omega_f$, therefore in the present case no resonant behavior should be expected.  
\begin{figure}
\begin{center}					  				
\rotatebox{0}{\scalebox{0.64}{\includegraphics{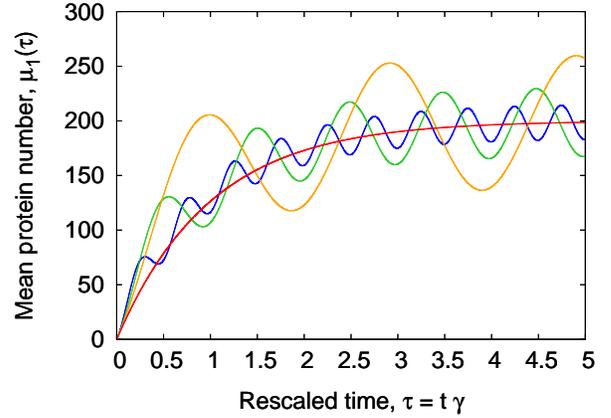}}} 
\end{center}  
\caption{(Color online) Mean molecule number $\kappa_1(\tau)= \mu_1(\tau)$ as given by Eq. (\ref{cumulant constant and periodic model parameters example sqrt}) for $r=1$, as a function of dimensionless time variable $\tau = \gamma t$, for $\kappa_1(0)= 0$, $\varphi=0$, $\gamma = 4 \cdot 10^{-4}$ s, $m_1$ and $m_2$ given by (\ref{moments 1st and 2nd of exponential nu}), $a = 10$, $b = 20$, $C_2 = a \gamma$,  $C_1 = C_2$ or $C_1 = 0$, and $T \gamma = \frac{1}{2}$ (blue), $T \gamma = 1$  (green), $T \gamma = 2$ (orange), and $C_1 = 0$ (time-independent $k(t)$, red curve).}
\label{Oscillating_mean}
\end{figure}
\begin{figure}
\begin{center}					  				
\rotatebox{270}{\scalebox{0.3}{\includegraphics{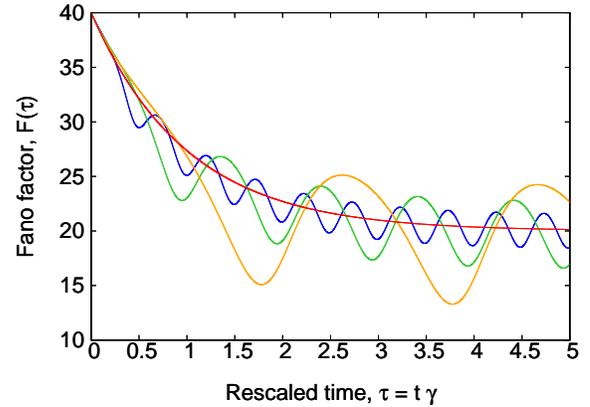}}} 
\end{center}  
\caption{(Color online) Fano factor $F(\tau) = \kappa_2(\tau)/\kappa_1(\tau)$ as a function of dimensionless time variable $\tau = \gamma t$, for $\kappa_1(0) = \kappa_2(0) = 0$, $\varphi=0$, 
$\gamma = 4 \cdot 10^{-4}$ s, $m_1$ and $m_2$ given by (\ref{moments 1st and 2nd of exponential nu}), $a = 10$, $b = 20$, $C_2 = a \gamma$,  $C_1 = C_2$ or $C_1 = 0$, and $T \gamma = \frac{1}{2}$ (blue), $T \gamma = 1$ (green), $T \gamma = 2$ (orange), and $C_1 = 0$ (time-independent $k(t)$, red curve).}
\label{Oscillating_Fano}
\end{figure}
In Fig. \ref{Oscillating_mean} we plot the time evolution of the average protein number $\kappa_1(\tau)= \mu_1(\tau)$ as a function of dimensionless time variable $\tau = \gamma t$ and for various oscillation frequencies corresponding to $T \gamma = \frac{1}{2}$ (blue), $T \gamma = 1$  (green) and $T \gamma  = 2$ (orange), as well as for the limiting case of nonoscillatory driving ($C_1 = 0$). We assume that $p(x,0)=\delta(x)$, therefore $\kappa_1(0) = \kappa_2(0) = 0$. Also, we assume here that $\nu(u)$ is an exponential distribution {(subscript $\epsilon$ stands for 'exponential')}
\begin{equation}
\nu_{\epsilon}(u) = \frac{1}{b}\exp\left(-\frac{u}{b} \right).
\label{exponential nu}
\end{equation}
Moments of $\nu_{\epsilon}(u)$ (\ref{exponential nu})  are given by 
\begin{equation}
m^{(\epsilon)}_n = b^n n!.
\label{moments of exponential nu}
\end{equation}
%
%
In particular, we have
\begin{equation}
m^{(\epsilon)}_1 = b,  ~~~~ m^{(\epsilon)}_2 = 2b^2.
\label{moments 1st and 2nd of exponential nu}
\end{equation}

Exponentially (or geometrically) distributed sizes of translational bursts have been observed in \textit{E. coli} \cite{cai2006stochastic, yu2006probing, taniguchi2010quantifying, choi2008stochastic}. For that reason, $\nu_{\epsilon}(u)$ (\ref{exponential nu}) appears to be a natural choice of the burst size distribution in the case of stochastic models of gene expression in which particle concentration is used instead of discrete particle number. {Note that} any other choice of $\nu(u)$ can only affect values of $m_1$ and $m_2$ in Eq. (\ref{cumulant constant and periodic model parameters example sqrt}); this results in  identical rescaling of each plot along the $y$-axis. As can be inferred from Eq. (\ref{cumulant constant and periodic model parameters example sqrt}), the amplitude of oscillation is the largest for the largest oscillation period.

Similarly, in Fig. \ref{Oscillating_Fano} we plot the time evolution of the Fano factor  $F(\tau)$ (\ref{Fano factor and eta}), again as a function of dimensionless time variable $\tau = \gamma t$ and for the same model parameters as in Fig \ref{Oscillating_mean}. By employing the L'H\^{o}pital's  rule, it can be shown that for $\kappa_1(0) = \kappa_2(0) = 0$ and $\nu(u)=\nu_{\epsilon}(u)$ (\ref{exponential nu}) we have 
\begin{equation}
\lim_{ \tau \to 0} F(\tau) = 2b,
\end{equation}
which is close to value ($F(0^{+}) = 2b+1$) obtained in Ref. \citep{thattai2001intrinsic} {for a similar discrete model}. 

Most of the results of the present Section can be immediately generalized to the case of arbitrary periodic dependence of burst frequency 
\begin{equation}
k(t) = a^{(k)}_{0} + \sum_{q=1}^{\infty} \left[ a^{(k)}_{q} \cos\left(q \omega_f t \right) + b^{(k)}_{q} \sin\left(q \omega_f t \right)\right].
\label{periodic k}
\end{equation}
Invoking (\ref{cumulant most general}), for $k(t)$ given by (\ref{periodic k}) we obtain
\begin{eqnarray}
\kappa_r(t) &=& \mathcal{T}_r(t) + \mathcal{P}_r(t) + \frac{a^{(k)}_{0}m_r}{r \gamma},   
\label{cumulant constant and periodic model parameters example general periodic k division into constant, periodic and transient parts}
\end{eqnarray}
%
where 

%
\begin{eqnarray}
\mathcal{T}_r(t) &=& \left(\kappa_r(0) + \sum_{q=0}^{\infty}  \frac{b^{(k)}_{q}  q \omega_f - a^{(k)}_{q} r \gamma}{r^2 \gamma^2 + q^2 \omega^2_f} m_r \right) e^{-r\gamma t}, \nonumber \\
\label{cumulant constant and periodic model parameters example general periodic k mathcal T}
\end{eqnarray}

and
\begin{eqnarray}
\mathcal{P}_r(t) &=& \sum_{q=1}^{\infty}  a^{(k)}_{q} \left( \frac{q \omega_f\sin(q \omega_f t)+ r \gamma \cos(q \omega_f t)}{r^2 \gamma^2 + q^2 \omega^2_f} \right) m_r \nonumber \\ &+&  \sum_{q=1}^{\infty}  b^{(k)}_{q} \left( \frac{r \gamma \sin(q \omega_f t)-q \omega_f \cos(q \omega_f t)}{r^2 \gamma^2 + q^2 \omega^2_f} \right) m_r. \nonumber \\
\label{cumulant constant and periodic model parameters example general periodic k mathcal P}
\end{eqnarray}
In Appendix \ref{Appendix A} we show that the division of $\kappa_r(t)$ into constant, transient and periodic part as given by (\ref{cumulant constant and periodic model parameters example general periodic k division into constant, periodic and transient parts}) remains valid when not only $k(t)$, but also $\gamma(t)$ or $\nu(u, t)$ are periodic functions of time.


Finally, let us note that Eq. (\ref{time evolution of r-th cumulant derived from K}) with $k(t)$ given by (\ref{sinusoidal k}) or, in general case, by (\ref{periodic k}) has a simple mechanical interpretation. Namely, it is the equation of motion of a particle moving with velocity $v=\kappa_r$ in a viscous medium under the influence of  both the drag force ($-r \gamma \kappa_r(t)$, with constant $\gamma$) and the external periodic force $m_r(t)k(t)$. Perhaps an even more compelling analogy is the RC low-pass filter: Fast oscillations of the external driving of gene expression \cite{mugler2010information} have less effect than slow ones.

\subsection{Time-independent model parameters\label{time-independent model parameters}}
\subsubsection*{Time evolution of $p(x,t)$}
When the model parameters do not depend on time, i.e., $k(t)=k$, $\gamma(t)=\gamma$, and $\hat{w}(s,t)=\hat{w}(s)$, Eq. (\ref{solution of the most general equation of characteristics}) may be rewritten as 
\begin{equation}
\hat{p}(s,t)=\Phi(s \Omega(t)) \exp\left[a\Psi(s)-a\Psi(s \Omega(t))\right],
\label{solution of the equation of characteristics for constant model parameters}
\end{equation}
%
%
%
where 
\begin{equation}
a=\frac{k}{\gamma},
\label{definition of a parameter}
\end{equation}
\begin{equation}
\Psi(z) = \int \frac{\hat{w}(z)}{z}dz.
\label{definition of Psi}
\end{equation}
$\Phi(z)$ is given again by Eq. (\ref{definition of Phi}), whereas
\begin{equation}
\Omega(t) = \exp(-\gamma t)
\label{definition of omega}
\end{equation}
is a special case of (\ref{definition of Omega and Gamma}). In the steady-state limit, from (\ref{solution of the equation of characteristics for constant model parameters}) we obtain
\begin{equation}
\lim_{t\to\infty}\hat{p}(s,t)\equiv \hat{p}(s) = \exp\left[a\left(\Psi(s)-\Psi(0)\right)\right].
\label{solution of the equation of characteristics for constant model parameters stationary}
\end{equation}
The form of stationary distribution $p(x)$ (we distinguish stationary and nonstationary probability distribution functions by the number of arguments) depends neither on the values of $k$ and $\gamma$ parameters alone, nor on the initial condition, but only on the functional form of the burst size pdf $\nu(u)$ and value of the parameter $a$ (\ref{definition of a parameter}).

Using (\ref{solution of the equation of characteristics for constant model parameters stationary}), we may rewrite (\ref{solution of the equation of characteristics for constant model parameters}) as
\begin{equation}
\hat{p}(s,t)=\Phi(s \Omega(t)) [\hat{p}(s\Omega(t))]^{-1} \hat{p}(s).
\label{solution of the equation of characteristics for constant model parameters product of three terms}
\end{equation}
Invoking the following property of Laplace transform \cite{abramowitz1964handbook}
\begin{equation}
\mathcal{L}^{-1}\left[\hat{f}(\alpha s)\right] = \frac{1}{\alpha}f\left(\frac{x}{\alpha}\right),
\label{scaling of laplace trasform}
\end{equation}
where $\hat{f}(s) = \mathcal{L}[f(x)]$, by taking the inverse Laplace transform of (\ref{solution of the equation of characteristics for constant model parameters product of three terms}) we can express $p(x, t)$ as the convolution of three terms 
\begin{equation}
p(x,t)= \frac{1}{\Omega(t)} p\left(\frac{x}{\Omega(t)},0\right) \ast p(x) \ast \frac{1}{\Omega(t)} q\left(\frac{x}{\Omega(t)}\right), 
\label{t dep prob as convolution}
\end{equation}
where 
\begin{equation}
p(x) = \mathcal{L}^{-1}\left[ \hat{p}(s) \right], ~~~~ q(x) = \mathcal{L}^{-1}\left[ 1/\hat{p}(s) \right].
\label{definition of q function}
\end{equation}
$\hat{p}(s)$ (\ref{solution of the equation of characteristics for constant model parameters stationary}) and $1/\hat{p}(s)$ cannot simultaneously satisfy the necessary conditions required for the Laplace transform of an ordinary function, in particular the condition $\lim_{s \to \infty}\hat{f}(s)=0$. Clearly, the latter condition should be obeyed by $\hat{p}(s)$, hence we have  $\lim_{s \to \infty} (1/\hat{p}(s))=\infty$. This implies that $q(x)$ (\ref{definition of q function}) is not an ordinary function, but a distribution consisting of ({apart from some ordinary function}) superposition of delta distribution and its derivatives. {In particular, if $1/\hat{p}(s)$ is a polynomial of degree $M$, 
\begin{equation}
\frac{1}{\hat{p}(s)}  = \sum_{k=0}^{M} q_k s^{k},
\label{1 over p in s space}
\end{equation}
we obtain} 
\begin{equation}
q(x) = \sum_{k=0}^{M} q_k \delta^{(k)}(x).
\label{explicit form of q function}
\end{equation}
If the explicit form of both $p(x)$ and $q(x)$  (\ref{explicit form of q function}) is known, it may be feasible to find the explicit form of $p(x,t)$ by invoking
Eq. (\ref{t dep prob as convolution}) and the identity 
\begin{equation}
(\delta^{(k)}\ast f)(x)=f^{(k)}(x).
\label{convolution of delta derivatives}
\end{equation}
%
%
Derivative on the r.h.s of Eq. (\ref{convolution of delta derivatives}) should be understood as a distribution derivative \cite{schwartz1965mathematical}. Namely, if $f(x)$ has a discontinuity at $x=0$, but is at least $m$ times differentiable for $x\neq 0$, the $m$-th distribution derivative of $f(x)$ reads 
\begin{equation}
f^{(m)} = \{f^{(m)}\} + \sigma_0 \delta^{(m-1)} + \sigma_1 \delta^{(m-2)} + \ldots + + \sigma_{m-1} \delta,
\label{distribution derivative}
\end{equation}
where $\{f^{(m)}\}$ denotes distribution related to $f^{(m)}$ treated as an ordinary function (not defined at $x=0$), whereas $\sigma_k = f^{(k)}(0^{+})-f^{(k)}(0^{-})$ \cite{schwartz1965mathematical}.  
In Appendix \ref{Appendix F} we apply Eqs. (\ref{solution of the equation of characteristics for constant model parameters product of three terms})-(\ref{distribution derivative}) to obtain solution of Eq. (\ref{unregulated t dependent ME of Friedman}) with the exponential probability distribution of burst sizes (\ref{exponential nu}) in an alternative way than the one used in Section \ref{Example exponential}.
\subsubsection*{Time evolution of cumulants of $p(x,t)$}
If the model parameters do not depend on time, Eq. (\ref{cumulant most general}) takes a remarkably simple form 
\begin{equation}
\kappa_r(t) = \kappa_r(0)e^{-r\gamma t} + a\left(1-e^{-r\gamma t}\right) \frac{m_r}{r},
\label{cumulant constant model parameters}
\end{equation}
where  $m_r$ is given by Eq. (\ref{moments of ni def}) with $\nu(u, t)=\nu(u)$. In the $t \to \infty $ limit, from (\ref{cumulant constant model parameters}) we obtain
\begin{equation}
\kappa_r = \kappa_r(\infty) = a \frac{m_r}{r}.
\label{cumulant constant model parameters stationary}
\end{equation}
which also follows from Eq. (\ref{how nu depends on p}) {of Appendix \ref{Appendix D} (in this Appendix, we further elaborate on the relationship between functional form of the burst size probability distribution $\nu(u)$ and the functional form of the corresponding steady-state distribution of protein concentration, $p(x)$).} Using (\ref{cumulant constant model parameters stationary}), we may rewrite (\ref{cumulant constant model parameters}) as
\begin{equation}
\kappa_r(t)=\kappa_r(0)e^{-r\gamma t} + \kappa_r(\infty)\left(1-e^{-r\gamma t}\right).
\label{cumulant constant model parameters more compact}
\end{equation}
The time evolution of $\kappa_r(t)$ as given by (\ref{cumulant constant model parameters}) or (\ref{cumulant constant model parameters more compact}) consists of the exponentially decaying contribution coming from the initial probability distribution $p(x,0)$, as well as the contribution proportional to the stationary distribution (\ref{cumulant constant model parameters stationary}); the latter is completely determined solely by the values of $a$ and $m_r$.

In the present case, if only the initial distribution $p(x,0)$ is known, the time evolution of cumulants may be immediately recovered from (\ref{cumulant constant model parameters}) if needed. This allows us to concentrate solely on the stationary limit ($t \to \infty$). Next, by making use of (\ref{ln pdf series expansion of Laplace transform gives cumulants}) and (\ref{cumulant constant model parameters}), we can obtain $\hat{p}(s,t)$ in the form of a power series in $s$ variable.

From Eqs. (\ref{cumulant constant model parameters}) or (\ref{cumulant constant model parameters more compact}) we see that the higher the cumulant order $r$ is, the faster $\kappa_r(t)$ approaches its stationary value. In particular, variance approaches stationary value faster than the mean protein concentration. 
For $r=1$, from (\ref{cumulant constant model parameters stationary}) we obtain a simple relation,
\begin{equation}
\kappa_1 = \mu_1 =  m_1 a.
\label{average protein number is a times average burst size}
\end{equation}
The parameter $a$ as defined by Eq. (\ref{definition of a parameter}) is equal to the burst frequency, $k$, multiplied by the characteristic time scale of the system, $T_{\gamma}=1/\gamma$. Therefore $a$ is proportional to the mean number of bursts (in Ref. \cite{friedman2006linking} parameter $a$ itself is called the burst frequency) and (\ref{average protein number is a times average burst size}) has a simple interpretation, i.e., the average protein concentration (number) is the average burst size times the mean number of bursts in time interval of the length $1/\gamma$.

For $r=2$, Eq. (\ref{cumulant constant model parameters}) can be rewritten as
\begin{equation}
\kappa_2(t) = \kappa_2(0)e^{-2\gamma t} + \frac{a}{2}\left(1-e^{-2\gamma t}\right) \left( \sigma^2(u) + b^2 \right),
\label{second cumulant constant model parameters division of noise}
\end{equation}
where $m_1=b$ and $\sigma^2_{\nu}(u)=m_2-b^2$ is the variance of $\nu(u)$. The term proportional to $\sigma_{\nu}^2(u)$ in Eq. (\ref{second cumulant constant model parameters division of noise}) is related to the stochasticity of the burst size distribution. However, even for the dispersionless ($\sigma_{\nu}(u)=0$) burst size distribution, 
\begin{equation}
\nu_{\delta}(u) = \delta\left(u-{b} \right),
\label{delta nu}
\end{equation}
%
%
we have an irreducible contribution to $\kappa_2(t)$ coming from the term proportional to $b^2$ in Eq. (\ref{second cumulant constant model parameters division of noise}). For a fixed $m_1=b$, $\nu_{\delta}(u)$ (\ref{delta nu}) minimizes the variance of $p(x,t)$, a result which could be intuitively expected.

\subsection{Example: p(x,t) corresponding to the exponential burst size distribution\label{Example exponential}}

In this section we find the time-dependent solution of Eq. (\ref{unregulated t dependent ME of Friedman}) for the exponential burst size distribution (\ref{exponential nu}).
Apart from gene expression models, exponential distribution (\ref{exponential nu}), as well as closely related two-sided exponential distribution found applications in models of other phenomena \cite{porporato2004phase,daly2006probabilistic,daly2010effect,suweis2011prescription,mau2014multiplicative}. It should be also noted that in most cases only for $\nu(u)$ of the form  (\ref{exponential nu}) both Eq. (\ref{unregulated t dependent ME of Friedman}) and its generalizations (e.g., jump-diffusion equations \cite{luczka1995white,czernik1997thermal,luczka1997symmetric}) are analytically tractable.

As shown in Ref. \cite{friedman2006linking}, for the time-independent model parameters, $\nu_{\epsilon}(u)$ (\ref{exponential nu}) leads to stationary distribution $p(x)$ in the form of gamma distribution, 
\begin{equation}
q_{\gamma}(x; a, b) \equiv \frac{x^{a-1} e^{-\frac{x}{b}}}{b^{a}\Gamma(a) } = \mathcal{L}^{-1}\left[\frac{1}{(sb+1)^a} \right].
\label{gamma distribution definition a b}
\end{equation}
This can be readily verified by making use of Eqs. (\ref{definition of Psi}) and (\ref{solution of the equation of characteristics for constant model parameters stationary}).

From  (\ref{moments of exponential nu}) and (\ref{cumulant constant model parameters}) we have 
\begin{equation}
\kappa^{(\epsilon)}_n(t) = \kappa_n(0)e^{-n\gamma t} + a\left(1-e^{-n\gamma t}\right) b^n (n-1)!
\label{cumulant constant model parameters epsilon}
\end{equation}
In the $t \to \infty$ limit, we obtain $\kappa^{(\epsilon)}_n = a b^n (n-1)!$, i.e., the cumulants of gamma distribution (\ref{gamma distribution definition a b}).
From (\ref{ln pdf series expansion of Laplace transform gives cumulants}) and (\ref{cumulant constant model parameters epsilon}) the Taylor series expansion of $\ln[\hat{p}_{\epsilon}(s)]$ can be reconstructed, we get
\begin{equation}
\ln[\hat{p}_{\epsilon}(s)] = a\sum_{n=1}^{\infty} \frac{(-bs)^n}{n} = \ln \left[\frac{1}{(sb+1)^a} \right],
\label{ln pdf series expansion of Laplace transform epsilon}
\end{equation}
hence $\hat{p}_{\epsilon}(s)=(sb+1)^{-a}$, which is indeed the inverse Laplace transform of gamma distribution (\ref{gamma distribution definition a b}).

Interestingly, in the present case both the explicit expression for $\hat{p}_{\epsilon}(s,t)$ and even for $p_{\epsilon}(x, t)$ can be obtained, at least for the initial distribution of the form
\begin{equation}
p_{\epsilon}(x,0)=\delta(x-x_0),
\label{initial condition x}
\end{equation}
where $x_0 \geq 0$ is the initial molecule concentration. The Laplace transform of (\ref{initial condition x}) is $\hat{p}_{\epsilon}(s,0)=\exp(-x_0 s)$, and hence from (\ref{solution of the equation of characteristics for constant model parameters}) we obtain
\begin{eqnarray}
\hat{p}_{\epsilon}(s,t) &=& \left( \frac{s e^{-\gamma t}+\frac{1}{b}}{s+\frac{1}{b}} \right)^{a} \exp(-x_0 e^{-\gamma t} s). 
\label{p hat in s space}
\end{eqnarray}
For simplicity, we put $x_0 = 0$ (which is arguably the most natural choice in the case of gene expression models). Moreover, we confine our attention to $a = n \in \mathbb{N}$, as only in this case we were able to find compact analytical expression for the inverse Laplace transform of $\hat{p}_{\epsilon}(s,t)$ (\ref{p hat in s space}). Still, (\ref{p hat in s space}) is valid for arbitrary real $a>0$. It is also convenient to change the independent variable according to $t \to \omega=\exp(-\gamma t)$. In such a case, $\tilde{p}_{\epsilon, n}(x,\omega(t)) \equiv   p_{\epsilon, n}(x,t) = \mathcal{L}^{-1}\{\hat{p}_{\epsilon, n}(s,t)\}$ reads
\begin{eqnarray}
\tilde{p}_{\epsilon, n}(x,\omega) &=& \omega^n \delta(x) + \sum_{i=1}^{n} {{n}\choose{i}}\frac{(1-\omega)^{i}\omega^{n-i}}{(i-1)!b^{i}}x^{i-1}e^{-\frac{x}{b}} \nonumber \\ &\equiv & \omega^n \delta(x) +  \sum_{i=1}^{n} {{n}\choose{i}}(1-\omega)^{i}\omega^{n-i} q_{\gamma}(x; i,b), \nonumber \\
\label{solution for p no hat x}
\end{eqnarray}
%
where $q_{\gamma}(x; i,b)$ is given by (\ref{gamma distribution definition a b}), whereas by $\hat{p}_{\epsilon, n}(s,t)$ we denote $\hat{p}_{\epsilon}(s,t)$ (\ref{p hat in s space}) for $a=n$ and similarly for $p_{\epsilon, n}(x,t)$ and $\tilde{p}_{\epsilon, n}(x,\omega)$.
Each of $\tilde{p}_{\epsilon, n}(x,\omega)$ functions (\ref{solution for p no hat x}) for $n=1, 2, \ldots$ is a superposition of gamma distributions (Dirac delta can be also treated as a limiting case of the gamma distribution) with different integer values of $a$ and time-dependent weights. Hence, (\ref{solution for p no hat x}) is a natural time-dependent generalization of the gamma distribution (\ref{gamma distribution definition a b}) with $a=n$, obtained in \cite{friedman2006linking}, where only the stationary limit of Eq. (\ref{unregulated t dependent ME of Friedman}) has been considered.

Note that for $x_0=0$, {the dependence of $\hat{p}_{\epsilon}(s,t)$ (\ref{p hat in s space}) on $s$ and the mean burst size $b$ is of the form} (\ref{scaling of Laplace transform of protein distribution by b}), therefore
$p_{\epsilon}(x,t)=\mathcal{L}[\hat{p}_{\epsilon}(s,t)]$, and in particular $p_{\epsilon, n}(x,t)=\tilde{p}_{\epsilon, n}(x,\omega(t))$ (\ref{solution for p no hat x}) have the characteristic dependence on $x$ variable and $b$ parameter as given by (\ref{scaling of protein distribution by b}), cf. Appendix \ref{Appendix E}.

An alternative way of obtaining $\tilde{p}_{\epsilon, n}(x,\omega)$ (\ref{solution for p no hat x}), its generalization for $x_0 > 0$ and its explicit form for small $n$ are discussed in Appendix \ref{Appendix F}.


\section{Discussion, biological insights}

The stochastic description of the simple system studied here shares a common feature   
with the corresponding  deterministic model: The time evolution of the average protein number predicted by the stochastic model is identical with the time evolution of the protein concentration obtained from deterministic equations of kinetics (see Appendix \ref{Appendix B}). On the other hand, the evolution of the $n$-th cumulant of the protein number distribution in time depends solely on the behavior of the $n$-th moment of the burst size distribution in time, but it does not depend on its other  moments. In consequence, the time evolution of the average molecule number is identical for all burst size distributions which have the same first moments, if only the remaining model parameters are identical. If additionally the time dependence of the second moments of the burst size distributions is identical, we obtain an identical time dependence of the coefficient of variation and the Fano factor of the protein number distributions, the two important measures of gene expression noise. And therefore, the predictions of   stochastic models with bursty molecule production are, to a large extent,  universal as they do not depend on  other details of the burst statistics. This may explain the success of gene expression models that commonly assume exponentially distributed burst sizes, despite the fact that the experimental evidence for this particular burst size distribution can be found in only a few papers \cite{cai2006stochastic, yu2006probing, taniguchi2010quantifying, choi2008stochastic}. (Note that a somewhat similar conclusion about an unexpected universality of coarse-grained models was drawn by Pedraza et al. \cite{pedraza2008effects} in regards to statistics of waiting times between mRNA bursts.)

It should also be noted that the effective bursty dynamics results from the approximation based on integrating out fast degrees of freedom. In order to check the  range of validity of this approximation,
the dynamics of the effective model (e.g. with protein but without mRNA, as considered here) should be compared with the dynamics of the full model including both slow (protein copy number) and fast (mRNA copy number) degrees of freedom. However, it is expected that predictions of the latter model are in agreement with the predictions of the former for $t$ greater than few mRNA lifetimes \cite{thattai2001intrinsic,shahrezaei2008analytical}. 

The Eq. (\ref{cumulant constant model parameters}) shows that the relaxation of the variance is twice faster than that of the mean (Fig. \ref{fig:noise} A). This has been shown previously for the model of gene expression where mRNA was explicitly taken into account and all reactions were Poissonian \cite{thattai2001intrinsic}. The same has been shown in ref. \cite{bar2006noise} (supplementary information therein), without referring to any particular reaction statistics. Eq. (\ref{cumulant constant model parameters}), on the other hand, links that result with the moments of an arbitrary distribution of protein bursts. Below, we will discuss these results in the context of evolutionary optimization of biological processes with respect to time-dependent noise intensity, and also we will relate the behavior of  Eq. (\ref{cumulant constant model parameters}) to experimentally measurable scaling relations between protein mean and variance. Although our model does not account for extrinsic noise nor feedback in gene regulation, our analysis may shed some light on understanding of the relation between protein number statistics and underlying burst statistics.
  
\subsection{Optimization of protein level detection with respect to noise is dependent on the assumed measure of noise}

Suppose that a cell population expresses a protein at a certain level in given environmental conditions, and then the conditions abruptly change, which results in a change in the expression level. How does the width of the protein distribution vary over time before it reaches a new steady state? Although the stationary behavior of noise in gene circuits has been widely studied, somewhat less studies have been devoted to transient behavior of noise (see e.g. \cite{thattai2001intrinsic,tabaka2008accurate,palani2012transient,dixon2016transient}).

The difference in relaxation time scales of the protein mean and variance may result in a nonmonotonic or, at least, nonlinear dependence of noise on time. It would be tempting to put forward a hypothesis that this feature may be exploited by evolution for optimization of some processes with respect to noise: For example, let the gene expression be reduced from an induced level to a basal level, and suppose that this reduction should trigger some other processes in the cell. For the trigger to be maximally precise (such that all cells can detect the decrease in protein concentration at almost the same time), its threshold should not necessarily be located  precisely at the basal expression level, but perhaps somewhere higher, where the noise is minimal.

We will show below, however, that such interpretations are dependent on the function assumed to quantify noise. We do not know what measure of noise does the biological evolution use -- that probably depends on the nature of a specific biological process. Coefficient of variation $\eta(t)$ (\ref{Fano factor and eta}) seems to be a relatively natural choice because it measures the ratio of distribution width to its mean, so it is a dimensionless quantity. However, Fano factor $F(t)$ (\ref{Fano factor and eta}) is also frequently used  in literature, a function that measures the ratio of variance to the mean, i.e. the deviation of the process from Poissonian statistics. On the other hand, in the context of detection of transition between two expression levels, an equally natural choice may be the fractional change of the distribution width between the initial and final (stationary) state. One can easily see that each of these quantities behaves differently.

\begin{figure}
\begin{center}					  				
\rotatebox{0}{\scalebox{1}{\includegraphics{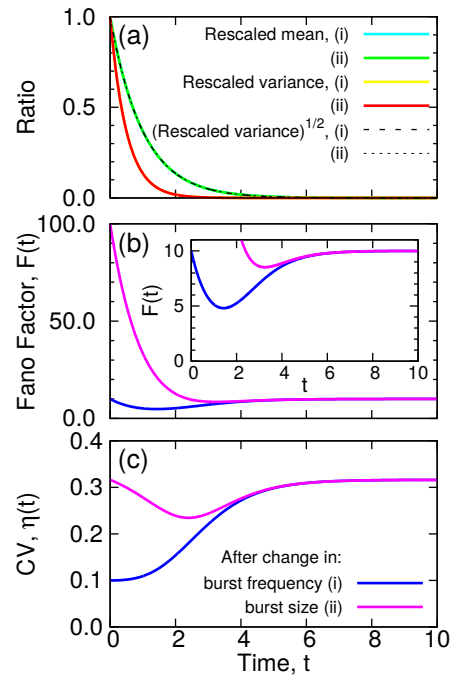}}}
\end{center}  
\caption{(Color online) Different measures of noise have a different transient behavior after an abrupt reduction of the mean burst frequency $a$ in a time unit defined by protein degradation (i), or the mean burst size $b$ (ii). The mean protein concentration $\langle x \rangle= ab$ was decreased from $10^3$ to $10^2$. A: Fractional change of mean, $K_1(t)$, fractional change of variance, $K_2(t)$, and its square root, $K_2(t)^{1/2}$. The curves for the cases (i) and (ii) overlap. B: Fano factor $F(t)$, inset: zoom to show the minima. C: Coefficient of variation, $\eta(t)$. }
\label{fig:noise}
\end{figure}

For visualization of the problem, suppose that the proteins are produced in exponential bursts of a mean size $b$. The number of proteins is therefore gamma-distributed with mean $ab$ and variance  $ab^2$. Let the initial expression level be $\langle x (0)\rangle = \kappa_1(0) = 10^3$ proteins, and after the abrupt environmental change it tends to $\langle x (\infty)\rangle = \kappa_1(\infty) = 10^2$. Such a change can be attained by two mechanisms: Decreasing $a$ (frequency modulation, FM) or decreasing $b$ (amplitude modulation, AM). Experimental evidence suggests that cells are able to adjust both $a$ and $b$ \cite{padovan2015single}. For $a=100$ and $b=10$, a ten-fold decrease in $\langle x \rangle$ by changing $a \rightarrow a/10$ at fixed $b$ results in a ten-fold change in variance (i). The same decrease in mean protein concentration by changing $b \rightarrow b/10$ at fixed $a$ yields a change in variance by the factor of $100$ (ii). 

In the case (i), the coefficient of variation has a deep minimum in $t=0$, and the Fano factor has a minimum at $t\neq 0$ (a relatively deep one, compared to the initial and final values). These dependencies are different in the case (ii): Here, the coefficient of variation has a shallow minimum at $t\neq 0$, and the Fano factor decreases almost monotonically by one order of magnitude, with a  minimum that is insignificant compared to the total change of $F(t)$ (see Fig. \ref{fig:noise} B, C).

The situation is still different if one takes into consideration the fractional change in the protein distribution width between the initial and final state. It immediately follows from Eq. (\ref{cumulant constant model parameters more compact}) that the fractional change of the $r$-th moment,
\begin{equation}\label{eq:fracch}
K_r(t) \equiv \frac{\kappa_r(t) - \kappa_r(\infty)}{\kappa_r(0) - \kappa_r(\infty)} = e^{-r \gamma t}.
\end{equation}
In particular, the square root of the fractional change of the variance is equal to the fractional change of the mean (Fig. \ref{fig:noise}). If $K_2(t)$ (or its increasing function) is used as the measure of the distribution width, then its minimal value is at $t=\infty$. And therefore, the optimization of the position of a detection threshold to minimize noise would be ambiguous, depending on a function chosen to quantify noise.

The above example shows that any biological hypotheses regarding the evolutionary optimization of some processes with respect to the amount of noise must assume that evolution has a specified method of measurement of that noise. If such optimizations really take place in nature, then it seems that the way how the evolution quantifies noise depends on a particular biological process. To date, it is not clear, however, which measure of noise is important in which process. This problem deserves a deepened experimental analysis.

\subsection{Frequency modulation and amplitude modulation cause different scalings of protein number variance to mean}

Experimental results suggest that cells can control gene expression levels both by adjusting the mean burst frequency $a$ (frequency modulation, FM) and the mean burst size $b$ (amplitude modulation, AM) \cite{padovan2015single}. With the Eq. (\ref{cumulant constant model parameters}) of our model, we can relate these two types of burst control to the scaling of mean and variance of protein distributions.

According to Eqs. (\ref{cumulant constant model parameters}) and (\ref{cumulant constant model parameters stationary}), the scaling of the variance of the protein number distribution with the mean depends on three parameters that describe burst statistics: mean burst frequency $a$, mean burst size $m_1$, and the second moment of the burst size distribution, $m_2$. For simplicity of notation, in the following discussion we will denote by $const.$ a constant whose value is universal for a set of different genes or for a single gene in cells cultured in various conditions. If the protein number distributions produced by the studied genes obey the scaling $\sigma^2_x / \langle x \rangle= \kappa_2(\infty) / \kappa_1(\infty) =F(\infty)=const.$ (i), then the moments of the burst size distribution depend on each other so that $m_2/m_1=const.$ and the mean burst frequency $a$ can be arbitrary. On the other hand, if one observes the scaling $\sigma^2_x / \langle x \rangle ^2  = \eta^2(\infty)=const.$ (ii), then the three burst parameters depend on each other so that $m_2/(a m_1^2)=const.$

If additionally the burst size distribution is such that $m_2/m_1^2 = \alpha = const.$, as in our examples in the main text and in the Appendix \ref{Appendix E} ($\alpha=1$ for delta burst size distribution, $\alpha=2$ for exponential, and $\alpha=(1+\lambda)/\lambda$ for gamma distribution, with $\lambda$ defined in the Eq. \ref{gamma nu}), then $\sigma^2_x /  \langle x \rangle=const.$ (i) implies that the mean burst size $m_1$ is universal, and the gene expression levels in the studied gene set, or in the set of conditions studied, are modulated by varying the mean burst frequency $a$ (FM). If, on the other hand, $\sigma^2_x /  \langle x \rangle^2=const$ (ii), then the mean burst frequency $a$ is universal and the gene expression level is modulated by mean burst size $m_1$ (AM). This dependence of the variance-to-mean relationship on AM or FM has been known \cite{hornung2012noise}, but an explicit or implicit assumption was that the burst size distributions are exponential. Here, we show that this property also extends to a class of nonexponential distributions.

Scaling (i) was observed, e.g., in \textit{S. cerevisiae} \cite{bar2006noise}, where different promoters controlled transcription of their native proteins fused with GFP, under different environmental conditions. Assuming burst size distributions such that $m_2/m_1^2 = \alpha$, would it be possible that the mean size of protein burst was the same in all these experiments  and only the burst frequency varied? This could perhaps be conceivable, if the protein burst size were globally limited by the availability of translational machinery, or if the mRNA of different GFP fusions had simultaneously similar stability and similar translation rates, such that the average number of proteins produced from one mRNA molecule was the same regardless of the gene. The burst frequencies could differ from gene to gene depending on the on/off switching rates of different promoters. 

However, we note that if the parameters of the burst size distribution are independent on time, then this fact imposes a particular form of scaling of the mean and variance of protein number distribution with time (\ref{cumulant constant model parameters}). In the Ref. \cite{bar2006noise}, the authors observed that when the variance and mean were normalized with respect to the initial state, $\Delta \kappa_r(t) \equiv [\kappa_r(t) - \kappa_r(0)] / \kappa_r(0)$, then the normalized variance was proportional to the normalized mean even in the time-dependent case out of the stationary state. Although the authors supposed that their theoretical model explained this scaling, it does not seem to be the case.  The equations for nonstationary mean and covariance proposed in \cite{bar2006noise} are fully consistent with the moment equations of our model and they imply that 
\begin{equation}
 \frac{\Delta \kappa_2(t)}{\Delta \kappa_1(t)} = \frac{ \Delta \kappa_2(\infty)}{\Delta \kappa_1(\infty)}(1+e^{-\gamma t}).
\end{equation}
The value in the braces changes from $2$ to $1$, so $\Delta \kappa_2(t)/ \Delta \kappa_1(t)$ cannot be maintained constant in time within our model, if the parameters of the burst distribution are constant in time. What if they are time-dependent? Using Eq. \ref{cumulant most general}, one can easily check that no simple substitution of exponential dependence of $\{m_1, \ m_2, \ k\}~\sim~\exp(-\gamma t)$ or $\sim~\exp(-2\gamma t)$ allows the function $\Delta \kappa_2(t)/ \Delta \kappa_1(t)$ to lose the dependence on time. The problem with time-dependent scaling suggests that as simple a model as ours may not be suitable for description of the data presented in \cite{bar2006noise}. This also suggests that more detailed studies  are needed on time dependence of protein noise and its relation to the properties of burst statistics.

Scaling (ii) was observed, e.g., in \textit{E. coli} and \textit{S. cerevisiae} cultured in different conditions \cite{salman2012universal}. The GFP gene was inserted under the control of three different promoters in multiple-copy plasmids (5 or 15 copies), or integrated into the genome in a single copy. If the distributions of burst sizes were such that $m_2/m_1^2 = \alpha$, would it be possible to explain this scaling behavior within our model? Gene expression level should be then modulated by the mean burst size, and the burst frequency should be universal. The quadratic scaling of mean vs. variance (\cite{salman2012universal}, Fig. 3 therein) can  be fitted by a one-parameter parabola $A\langle x \rangle ^2$. We note that the values of $A$ are different for the three different promoters. Within our model, this would mean that each promoter has its own characteristic frequency of bursting. This would sound reasonable if single gene copies were studied. However, in the experiments of \cite{salman2012universal}, the promoters were present in variable numbers of copies. The burst frequencies of the  gene copies should then add up (see Eq. \ref{sum of k i equal k}) and a single promoter should have a lower burst frequency than its multiple copies, unless there is some mechanism of dosage compensation in cells, which keeps the total burst frequency independent on the copy number of a given promoter. Moreover, the universal scaling of the full protein number distributions in \cite{salman2012universal} was defined by a function $\varphi((x-\langle x \rangle)/\sigma)$ (where $\sigma$ denotes  standard deviation), so, for example, gamma distribution produced by exponential bursts does not obey that scaling. Therefore, the validity of our model with time-independent parameters and the AM modulation of gene expression seems unlikely in the case of the data presented in \cite{salman2012universal}. 

Yet, the above considerations based on our simple theory reveal that there are still unexplored problems in the field of stochastic gene expression: Are the distributions of protein  burst sizes constant in time? Do they always belong to the wide class of those fulfilling $m_2/m_1^2=\alpha$, which includes the exponential distribution, commonly assumed in modeling? Under what biological conditions are the protein number distributions controlled by amplitude modulation of protein bursts or by frequency modulation (AM vs. FM)? Do these mechanisms undergo dosage compensation in the case of gene multiplication? The present discussion may be, therefore, an inspiration for a deepened experimental analysis of time dependence of protein number statistics on the underlying burst size statistics.

\section*{Acknowledgments}
The research was partly supported by the Ministry of Science and Higher Education Grant No. 0501/IP1/2013/72 (Iuventus Plus).

\vspace{0.3cm}
$\ast$Corresponding author, e-mail: jjedrak@ichf.edu.pl

\appendix

{ 
\section{Deterministic rate equation \label{Appendix B}}
The deterministic model of the reaction kinetics for the system described by Eq. (\ref{biochemical reaction scheme}) is given by the following rate equation}
\begin{eqnarray}
\dot{x} &=& c(t) - \gamma(t) x, 
\label{general deterministic ode}
\end{eqnarray}
where $x \geq 0$ is the molecule concentration, $c(t)$ is the source intensity,  whereas $\gamma(t)$ is the decay parameter; dot denotes the time derivative. {Note that $c(t)$ describes a deterministic birth process, in contrast to the random source intensity $I(t)$ of the corresponding stochastic model (\ref{langevin ode}).} The solution to the Eq. (\ref{general deterministic ode}) for $x(0)=x_0$ can be readily obtained: 
\begin{equation}
x(t)=e^{-\int_{t_0}^{t} \gamma(t^{\prime}) dt^{\prime}}\left[x_0 + \int_{t_0}^{t} c(t^{\prime})   e^{\int_{t_0}^{t^{\prime}} \gamma(\tilde{t}) d\tilde{t}} d t^{\prime} \right].
\label{solution of general deterministic ode}
\end{equation}
The existence of the stationary solution to Eq. (\ref{general deterministic ode}), as well as the possible oscillatory character of molecule concentration $x(t)$ depends on the functional form of $c(t)$ and $\gamma(t)$. Clearly, in the case of time independent model parameters, i.e., $c(t)=c$ and $\gamma(t)=\gamma$, the unique, stable stationary point of (\ref{general deterministic ode}) is given by
\begin{equation}
x_s = \frac{c}{\gamma}.
\label{deterministic stationary point}
\end{equation}

\section{Time-evolution of the moments of p(x,t)\label{Appendix C}}
Multiplying Eq. (\ref{unregulated t dependent ME of Friedman}) by $x^r$ and integrating such obtained equation, one gets the time evolution equation for the $r$-th moment of $p(x, t)$, 
\begin{equation}
\mu_r(t) = \int_{0}^{\infty} x^r p(x, t) dx.
\label{moments of p def}
\end{equation}
In the resulting time evolution equation for $\mu_r(t)$, the term derived from the first term on the r.h.s of Eq. (\ref{unregulated t dependent ME of Friedman}) is readily integrated by parts, whereas in the term containing $\nu(x-x^{\prime}, t)$ we have to change order of integration with respect to $x$ and $x^{\prime}$ as well as to change the independent variables according to $(x, x^{\prime}) \to (u, x^{\prime})$, $u = x-x^{\prime}$, cf. Ref. \cite{feng2012analytical}. In result, we get
\begin{equation}
\dot{\mu}_r(t) = -r\gamma(t)\mu_r(t) + k(t)\sum_{q=1}^{r}\binom{r}{q}\mu_{r-q}(t)m_q(t),
\label{time evolution equation for moments of p}
\end{equation}
where $m_r(t)$ denotes $r$-th moment of the burst size distribution $\nu(u, t)$ {as given by Eq. (\ref{moments of ni def})}, i.e., 
\begin{equation}
m_r(t) = \int_{0}^{\infty} u^r \nu(u, t) du.
\label{moments of ni def appendix}
\end{equation}
We assume here that the burst size pdf $\nu(u, t)$ is properly normalized and that the normalization is conserved during time evolution, i.e. $m_0(t)=1$, but we impose no other restrictions on the functional form of $\nu(u, t)$. On the other hand, the normalization of $p(x, t)$, i.e., the condition $\mu_0(t)=1$ follows immediately from Eq. (\ref{time evolution equation for moments of p}). For $r=1$, Eq. (\ref{time evolution equation for moments of p}) reads    
\begin{equation}
\dot{\mu}_1(t) = k(t)m_1(t) - \gamma(t)\mu_1(t).
\label{time evolution equation for the first moment of p appendix}
\end{equation}
Eq. (\ref{time evolution equation for the first moment of p appendix}) is identical to Eq. (\ref{general deterministic ode}) provided that we put $c(t)=k(t)m_1(t)$. In such a case, the time evolution of the average molecule number $\mu_1(t)$ is the same as the time evolution of molecule concentration $x(t)$ in the corresponding deterministic model (\ref{general deterministic ode}); this is a general property of linear deterministic dynamical systems \cite{van2007stochastic, mcquarrie1967stochastic}. 
Therefore, making use of Eq. (\ref{solution of general deterministic ode}), we may immediately write down the solution of Eq. (\ref{time evolution equation for the first moment of p appendix})
\begin{equation}
\mu_1(t)=\Omega(t)\left[\mu_1(0) + \int_{t_0}^{t} \frac{k(t^{\prime})m_1(t^{\prime})}{\Omega(t^{\prime})}dt^{\prime} \right],
\label{solution of time evolution equation for the first moment of p appendix}
\end{equation}
where $\Omega(t)$ is defined by Eq. (\ref{definition of Omega and Gamma}), i.e., 
\begin{equation}
\Omega(t) = \exp \left(-\int_{t_0}^{t} \gamma(t^{\prime}) dt^{\prime}\right).
\label{definition of Omega and Gamma appendix}
\end{equation}
For each $n\in \mathbb{N}$, Eqs. (\ref{time evolution equation for moments of p}) with $r=1, 2, \ldots, n$ form closed hierarchy of linear differential equations (time evolution equation for $\mu_r(t)$ does not depend on $\mu_i(t)$ if $i>r$). Therefore, in principle, starting from $r=1$, the explicit analytical formula for  $\mu_n(t)$ can be found iteratively for arbitrary $n$.


\section{Periodic time dependence of the model parameters \label{Appendix A}}
Below, we show that the possibility of division of $\kappa_r(t)$ into two parts, the exponentially decaying and periodic one, as demonstrated on the simple example analyzed in Subsection \ref{oscillatory time dependence}, is in fact a general feature of the present model when its parameters depend periodically on time.  We assume now that not only $k(t)$ (\ref{periodic k}), but also $\gamma(t)$, and $m_r(t)$ appearing in (\ref{cumulant most general}) are periodic functions of time (including constant function as the special case), 
\begin{eqnarray}
\gamma(t) &=& \sum_{q=0}^{\infty} \left[ a^{(\gamma)}_{q} \cos\left(q \omega_f t \right) + b^{(\gamma)}_{q} \sin\left(q \omega_f t \right)\right],  \nonumber \\ 
\label{fourier series gamma}
\end{eqnarray}
\begin{eqnarray}
m_{n}(t) &=& \sum_{q=0}^{\infty} \left[ a^{(m)}_{n,q} \cos\left(q \omega_f t \right) + b^{(m)}_{n,q} \sin\left(q \omega_f t \right)\right], \nonumber \\ 
\label{fourier series m n}
\end{eqnarray}
where $\omega_f$ is given by Eq. (\ref{definition of omega f}). We also assume that at least one of $k(t)$, $\gamma(t)$,  and $m_{n}(t)$ functions is nontrivial periodic  function (i.e., is not a constant, $k(t) \gamma(t) m_{n}(t) \neq \text{const}$). From (\ref{fourier series gamma}) we obtain 
\begin{eqnarray}
\int_{t_0}^{t} \gamma(t^{\prime}) d t^{\prime} &=& \sum_{q=1}^{\infty} \frac{1}{q\omega_f}\left[ a^{(\gamma)}_{q} \sin\left(q \omega_f t \right) -  b^{(\gamma)}_{q} \cos\left(q \omega_f t \right)\right] \nonumber \\ &-&  \mathcal{A}^{(\gamma)}_0 + a^{(\gamma)}_{0}t,  
\label{integral of fourier series gamma}
\end{eqnarray}
where
\begin{eqnarray}
\mathcal{A}^{(\gamma)}_0 &=& a^{(\gamma)}_{0}t_0 + \sum_{q=1}^{\infty} \frac{1}{q \omega_f }  a^{(\gamma)}_{q} \sin\left(q \omega_f t_0 \right) \nonumber \\ &-& \sum_{q=1}^{\infty} \frac{1}{q \omega_f} b^{(\gamma)}_{q} \cos\left(q \omega_f t_0\right).
\label{integral of fourier series gamma definition of mathcal A 0}
\end{eqnarray}
%
From (\ref{integral of fourier series gamma}) and (\ref{integral of fourier series gamma definition of mathcal A 0}) it follows that $[\Omega(t)]^{-r}$ can be written as 
\begin{equation}
[\Omega(t)]^{-r} =  e^{r a^{(\gamma)}_{0}t } \mathcal{P}_1(t)
\label{Omega divided into exponentially decaying and periodic parts Appendix}
\end{equation}
where $\mathcal{P}_1(t)$ is a periodic function of time and  $\Omega(t)$ is given by (\ref{definition of Omega and Gamma})  or (\ref{definition of Omega and Gamma appendix}). From (\ref{Omega divided into exponentially decaying and periodic parts Appendix}) it follows that the integrand appearing in Eq. (\ref{cumulant most general}), i.e., $k(t) m_r(t)/ \left[\Omega(t)\right]^{r}$ is also of the form (\ref{Omega divided into exponentially decaying and periodic parts Appendix}) but with $\mathcal{P}_1(t)$ replaced by another periodic function, $\mathcal{P}_2(t) = \mathcal{P}_1(t) k(t) m_r(t)$ (product of finite number of periodic functions is a periodic function itself). {Next}, consider the integral 
\begin{eqnarray}
\mathcal{I} &=& \int_{t_0}^{t}  e^{r a^{(\gamma)}_{0} t^{\prime}}  \mathcal{P}_2(t^{\prime}) dt^{\prime}.
\label{mathcal I is an integral with mathcal P 2}
\end{eqnarray}
Because $\mathcal{P}_2(t^{\prime})$ can be expanded in a Fourier series, we are left with integrals of the form 
\begin{eqnarray}
\mathcal{I}_s(q) &=& \int_{t_0}^{t}  e^{r a^{(\gamma)}_{0} t^{\prime}}  \sin(q \omega_f t^{\prime}) dt^{\prime}.
\label{integral exp and sin}
\end{eqnarray}
\begin{eqnarray}
\mathcal{I}_c(q) &=& \int_{t_0}^{t}  e^{r a^{(\gamma)}_{0} t^{\prime}}  \cos(q \omega_f t^{\prime}) dt^{\prime}.
\label{integral exp and cos}
\end{eqnarray}
The integrals (\ref{integral exp and sin}) and (\ref{integral exp and cos}) are elementary, we have
\begin{equation}
\int e^{\lambda t}  \sin(\alpha t) dt = \frac{\lambda\sin(\alpha t) - \alpha \cos(\alpha t)}{\alpha^2 + \lambda^2}e^{\lambda t}
\label{integral exp and sin indefinite explicit}
\end{equation}
and
\begin{equation}
\int e^{\lambda t}  \cos(\alpha t) dt = \frac{\lambda\cos(\alpha t) + \alpha \sin(\alpha t)}{\alpha^2 + \lambda^2}e^{\lambda t},
\label{integral exp and cos indefinite explicit}
\end{equation}
i.e., indefinite integral of the type (\ref{integral exp and sin}) or (\ref{integral exp and cos}) is a product of the exponential function and the periodic part containing $\sin(\alpha t)$ and $\cos(\alpha t)$ terms; definite integrals (\ref{integral exp and sin}) and (\ref{integral exp and cos}) contain also the constant (time-independent) part.

In result, the exponential term $e^{-r a^{(\gamma)}_{0} t}$ coming from $\left[\Omega(t)\right]^{r}$ in front of the integral in  Eq. (\ref{cumulant most general}) cancels the corresponding exponential term $e^{r a^{(\gamma)}_{0} t}$ appearing in some (but not all) terms of definite integral (\ref{mathcal I is an integral with mathcal P 2}); such terms depend on $t$ in a periodic manner. On the other hand, the remaining terms in  Eq. (\ref{cumulant most general}) are not periodic, but exponentially decaying functions of time.  
\vspace{0.3cm}

\section{Infinite divisibility property of protein concentration probability distribution  \label{Appendix D}}
Equations (\ref{definition of Psi}) and (\ref{solution of the equation of characteristics for constant model parameters stationary}) establish a mapping between $\hat{p}(s)=\hat{p}(s; a)$ and $\hat{\nu}(s)$, or, equivalently, between $p(x; a)$ and $\nu(u)$ (here we explicitly denote the dependence of probability distributions on the parameter $a$ both in $x$ and $s$ space). Given a particular stationary distribution of molecule concentration $p(x; a)$, we may ask what is the functional form of $\nu(u)$ for which $p(x; a)$ is a stationary solution of Eq. (\ref{unregulated t dependent ME of Friedman}). From Eq. (\ref{solution of the equation of characteristics for constant model parameters}) we obtain  
\begin{eqnarray}
\hat{\nu}(s) &=& \frac{s}{a} \frac{\hat{p}^{\prime}(s; a)}{\hat{p}(s; a)} + 1 = \frac{s}{a} \left(\ln[\hat{p}(s; a)]\right)^{\prime} + 1,
\label{how nu depends on p}
\end{eqnarray}
where prime denotes derivative with respect to $s$. 
Because $\hat{\nu}(s)$ is the Laplace transform of probability density function, for real $s\geq 0$ it must fulfill the following conditions: (i) $\hat{\nu}(0)=1$, (ii) $\hat{\nu}(s) \geq 0$ (iii) $\hat{\nu}^{\prime}(s)<0$, and (iv) $\lim_{s \to \infty }\hat{\nu}(s) = 0$. This imposes restrictions on the allowed form of $\hat{p}(s; a)$; otherwise the solution of Eq. (\ref{how nu depends on p}) satisfying (i)-(iv) may not exist. The condition (i) is fulfilled by arbitrary $\hat{p}(s; a)$, but conditions (ii), (iii) and (iv) involve parameter $a$. 
However, the dependence of any stationary solution $\hat{p}(s; a)$ of Eq. (\ref{unregulated t dependent ME of Friedman in s space}) on $a$ is of a special form; from (\ref{solution of the equation of characteristics for constant model parameters stationary}) we have 
\begin{equation}
\hat{p}(s; a)  = \left[\hat{p}(s; 1)\right]^a, 
\label{solution of the equation of characteristics for constant model parameters stationary how it depends on a}
\end{equation}
%
%
where 
\begin{equation}
\hat{p}(s; 1)=\exp[\Psi(s)-\Psi(0)]
\label{definition of f(s)}
\end{equation}
does not depend on $a$. Invoking (\ref{solution of the equation of characteristics for constant model parameters stationary how it depends on a}) and (\ref{definition of f(s)}), Eq. (\ref{how nu depends on p}) can be rewritten as
\begin{equation}
\hat{\nu}(s) = s \frac{\hat{p}^{\prime}(s; 1)}{\hat{p}(s; 1)} + 1.
\label{how nu depends on f}
\end{equation}
The special form (\ref{solution of the equation of characteristics for constant model parameters stationary how it depends on a}) of $\hat{p}(s; a)$ is a manifestation of the infinite divisibility property of stationary solutions of the present model. Equation (\ref{solution of the equation of characteristics for constant model parameters stationary how it depends on a}) can be inferred even without solving Eq. (\ref{unregulated t dependent ME of Friedman in s space}). Namely, within the present model, a single gene with burst frequency $k$ is equivalent to (and can be replaced with) $N$ parallel, independent gene copies (sources) characterized by burst frequencies $k_1, k_2, \ldots, k_N$, provided that
\begin{equation}
k_1 + k_2 + \ldots + k_N =k, 
\label{sum of k i equal k}
\end{equation}
(i.e., under the assumption that there is no additional mechanism of dosage compensation in cells which would decrease the burst frequency as the gene copy number increases) and that the burst size distributions are identical for each source, 
\begin{equation}
\nu_1(u)=\nu_2(u)=\ldots=\nu_N(u)=\nu(u). 
\label{all nu i are identical}
\end{equation}
Each of these $N$ independent gene copies alone generates probability distribution $p_i(x_i; a_i)$ of a random variable $x_i$, $i=1, 2, \ldots, N$, where $a_i= k_i/\gamma$, with $\gamma$ being a degradation constant, common for all molecules present in the system. We also have $a=a_1+a_2 + \dots + a_N$. 
If all $N$ gene copies are simultaneously present, the total molecule concentration, $x=x_1+x_2 + \dots + x_N$ is a sum of $N$  independent random variables $x_i$, and hence $p(x; a)$ is a convolution of $p_1(x_1; a_1), p_2(x_2; a_2), \ldots, p_N(x_N; a_N)$,
\begin{equation}
p(x; a) = p_1(x; a_1) \ast  p_2(x; a_2) \ast \cdots \ast p_N(x; a_N).
\end{equation}
In consequence, we have 
\begin{equation}
\hat{p}(s; a)=\mathcal{L}[p(x; a)]=\hat{p}_1(s; a_1)\hat{p}_2(s; a_1)\cdots  \hat{p}_N(s; a_1).
\label{hat p is a product of hat p i}
\end{equation}
In particular, for $a_1 = a_2 = \ldots = a_N =a/N$ we have $\hat{p}_1(s; a/N)=\hat{p}_2(s; a/N)=\hat{p}_N(s; a/N)$, therefore
%
%
\begin{equation}
\hat{p}(s; a)=\hat{p}(s; a/N)^{N}. 
\label{functional equation for f}
\end{equation}
Clearly, $\hat{p}(s; a)$ (\ref{solution of the equation of characteristics for constant model parameters stationary how it depends on a}) is the solution of the above functional equation. 

For the initial condition of the form $p(x, t_0)=\delta(x)$ ($\Phi(z)=1$), probability distribution of the molecule concentration exhibits the infinite divisibility property not only in the $t \to \infty$ limit, but for arbitrary $t\geq t_0$ as well. Indeed, in such a case (\ref{solution of the equation of characteristics for constant model parameters product of three terms}) reads
\begin{equation}
\hat{p}(s,t)=\frac{\hat{p}(s; a)}{\hat{p}(s\Omega(t); a)}  = \left(\frac{\hat{p}(s; 1)}{\hat{p}(s\Omega(t); 1)}  \right)^{a}.
\label{solution of the equation of characteristics for constant model parameters product of three terms with trivial initial condition}
\end{equation}
From Eq. (\ref{solution of the equation of characteristics for constant model parameters product of three terms with trivial initial condition}) we see that for $a=n \in \mathbb{N}$, $p(x, t; n)$ can be obtained as $n$-th convolution of $p(x, t; 1)$, 
\begin{equation}
p(x, t; n)=\underbrace{p(x, t; 1)\ast \cdots \ast p(x, t; 1)}_{\text{$n$~times.}}
\label{t dep prob in x as the n th convolution}
\end{equation}
From the above it follows that the property of infinite divisibility also works for the time-dependent distributions generated by our model, as long as the initial condition {is given by} delta function.
 
\section{Examples of burst size probability distributions \label{Appendix E}}
Below we analyze some properties of the solutions of Eq. (\ref{unregulated t dependent ME of Friedman}), obtained for two selected choices of the burst size pdf $\nu(u)$.

First, we analyze an example of a simple burst size pdf in the form of Dirac delta distribution. In this case, as well as for all burst pdfs of the form $\nu(u)=\nu(u; b)$, i.e.,  depending on a single parameter $b$, equal to mean burst size, we obtain a two-parameter family of probability distributions of molecule concentration $p(x, t)=p(x, t; a, b)$. Next, we analyze the three-parameter family of probability distributions $p(x,t)=p(x,t;a, \lambda, \xi)$ obtained for the burst size distribution given by gamma distribution with parameters $\lambda$ and $\xi$. In order to compare this case with the remaining ones (Dirac delta and exponential distribution analysed in Section \ref{Example exponential}), we put $\lambda \xi = b$. Therefore for all burst size distributions considered {in the present paper}, all moments $m_n$, $n = 1, 2, 3, \dots $ are finite and $m_1=b$. Also, in each case {analyzed here}, the dependence of burst size probability distribution on burst size $u$ and mean burst size $b$ is of the form 
\begin{equation}
\nu(u; b)=\frac{1}{b}h\left(\frac{u}{b}\right),
\label{scaling of burst distribution by b}
\end{equation}
where $h(y) \geq 0$ and $\int_{0}^{\infty}h(y)dy = 1$.
For any $\nu(u; b)$ of the form (\ref{scaling of burst distribution by b}), from (\ref{scaling of laplace trasform}) it follows that 
\begin{equation}
\hat{\nu}(s; b)=\mathcal{L}\left[\nu(u; b) \right]= \hat{h}(sb)
\label{scaling of Laplace transform of burst distribution by b}
\end{equation}
where $\hat{h}(s)=\mathcal{L}[h(u)]$. From (\ref{scaling of Laplace transform of burst distribution by b}), (\ref{definition of Psi}), (\ref{solution of the equation of characteristics for constant model parameters stationary}), and (\ref{solution of the equation of characteristics for constant model parameters product of three terms}) it follows, that for $\Phi(s)=1$, we also have
\begin{equation}
\hat{p}(s, t; b, \ldots) = \hat{g}\left(sb, t; \ldots  \right).
\label{scaling of Laplace transform of protein distribution by b}
\end{equation}
Therefore, for the initial distribution $p(x, t_0)=\delta(x)$, the property (\ref{scaling of burst distribution by b}) is inherited by $p(x, t)$, i.e., the dependence of the latter distribution on parameter $b$ is of the form  
\begin{equation}
p(x, t; b, \ldots)=\frac{1}{b}g\left(\frac{x}{b}, t; \ldots \right),
\label{scaling of protein distribution by b}
\end{equation}
where by $\ldots$ we denote the remaining model parameters. {Obviously, from (\ref{scaling of protein distribution by b}) it follows that the corresponding stationary distribution of protein concentration depends on $b$ in the same manner}, 
\begin{equation}
p(x; b, \ldots)=\frac{1}{b}g\left(\frac{x}{b}; \ldots \right).
\label{scaling of protein distribution by b stationary}
\end{equation}

\subsection{Dirac delta distribution}
The simplest form of the burst-size probability distribution $\nu(u)$ is the dispersionless delta distribution $\nu_{\delta}(u)$, Eq. (\ref{delta nu}), which describes identical bursts of size $b$. For this particular burst-size probability distribution, only the random distribution of the burst arrival times contributes 
to the stochastic character of the protein production, but there is no contribution of the burst-size fluctuations. Obviously, we have
\begin{equation}
m^{(\delta)}_n = b^n. 
\label{moments of delta nu}
\end{equation}
From (\ref{cumulant constant model parameters}), (\ref{cumulant constant model parameters stationary}), and (\ref{moments of delta nu}) we obtain now 
\begin{equation}
\kappa^{(\delta)}_n(t) = \kappa_n(0)e^{-n\gamma t} + a\left(1-e^{-n\gamma t}\right)\frac{b^n}{n},
\label{cumulant constant model parameters delta}
\end{equation}
\begin{equation}
\kappa^{(\delta)}_n(\infty) = a\frac{b^n}{n}.
\label{cumulant constant model parameters delta stationary limit}
\end{equation}
Equations (\ref{cumulant constant model parameters delta stationary limit}) and (\ref{ln pdf series expansion of Laplace transform gives cumulants}) allow us to find Taylor series expansion of $\ln[\hat{p}_{\delta}(s)]$, 
\begin{equation}
\ln[\hat{p}_{\delta}(s)] = a\sum_{n=1}^{\infty} \frac{(-bs)^n}{ n!n} = - a \text{Ein}(-bs),
\label{ln pdf series expansion of Laplace transform delta}
\end{equation}
where 
\begin{eqnarray}
\text{Ein}(z) &=& \int_{0}^{z}\left(1-e^{-t}\right)\frac{dt}{t} =\sum_{k=1}^{\infty}\frac{(-1)^{k+1} z^k}{k~ k!}
\label{definition of Ein function}
\end{eqnarray}
is related to exponential integral $\text{Ei}(z)$ in the following way \cite{abramowitz1964handbook}
\begin{equation}
\text{Ein}(-y) = -\text{Ei}(y) + \ln(y) + C_{\gamma}, ~~~~ y>0.
\label{definition of Ein}
\end{equation}
In Eq. (\ref{definition of Ein}), by $C_{\gamma}$  we denote the Euler-Mascheroni constant, usually denoted by $\gamma$.

\subsection{Gamma distribution}
{As a second example}, we consider the case of burst size pdf $\nu(u)$ given by gamma distribution 
\begin{equation}
\nu_{\gamma}(u;\lambda, \xi) = \frac{u^{\lambda-1}}{\xi^{\lambda} \Gamma(\lambda)}\exp\left(-\frac{u}{\xi} \right).
\label{gamma nu}
\end{equation}
$\nu_{\gamma}(u;\lambda, \xi)$ (\ref{gamma nu}) is one of the simplest continuous and differentiable unimodal burst size distribution functions. Moreover, burst size distributions similar to (\ref{gamma nu}) appear naturally in certain models of gene expression \cite{schwabe2012transcription}. This burst size distribution has also been analyzed in Ref. \cite{daly2010effect}. In order to compare $\nu_{\gamma}(u)$ with  burst size distribution analyzed above,    
we put $\lambda \xi = m^{(\gamma)}_1  = b$. Using the formula for $n$-th moment of gamma distribution 
%
\begin{equation}
m^{(\gamma)}_n = \frac{\Gamma(n+\lambda)}{\Gamma(\lambda)}\xi^{n}=\frac{\Gamma(n+\lambda)}{\lambda^{n}\Gamma(\lambda)}{b}^{n}, 
\label{moments of gamma nu}
\end{equation}
and Eq. (\ref{cumulant constant model parameters stationary}) we obtain 
%
\begin{equation}
\kappa^{(\gamma)}_n = a\frac{\Gamma(n+\lambda)}{n \Gamma(\lambda)}\xi^{n}.
\label{cumulant constant model parameters gamma stationary limit}
\end{equation}
From (\ref{ln pdf series expansion of Laplace transform gives cumulants}) and (\ref{cumulant constant model parameters gamma stationary limit}) we obtain 
\begin{equation}
\ln[\hat{p}_{\gamma}(s)] = a\sum_{n=1}^{\infty}  \frac{\Gamma(n+\lambda)}{\Gamma(\lambda) \Gamma(n+1)} \frac{(-\xi s)^n}{n}.
\label{ln pdf series expansion of Laplace transform gamma}
\end{equation}
%
Full time evolution of $\hat{p}_{\gamma}(s, t)$ can be easily recovered, if necessary. 
Further simplification are possible for $\lambda \in \mathbb{N}$. In such a case, instead of $s$ it is more convenient to use $\chi(s)=(sb+1)^{-1}$ as an independent variable. Because $\mathcal{L}^{-1}[(sb+1)^{-a}] = q_{\gamma}(x; a, b)$, cf. Eq. (\ref{gamma distribution definition a b}), the expansion of $\hat{p}_{\gamma}(s(\chi))$ in powers $\chi$ is equivalent with expressing $p(x)$ as a superposition of gamma distributions (\ref{gamma distribution definition a b}) with different values of $a$ but the same $b$. This 'gamma representation' may be viewed in analogy with widely used Poisson representation \cite{gardiner2009stochastic}.

In particular, for $\lambda=2$ the explicit formula for $p(x)$ can be obtained, we get
\begin{eqnarray}
p_{\gamma}(x;2, \xi, a)= \frac{1}{(e a)^a}\frac{e^{-\frac{x}{\xi}}}{x} \mathcal{F}_a\left(\frac{x}{\xi}\right),
\label{p(x) for gamma bursts and lambda equal 2}
\end{eqnarray}
where
\begin{eqnarray}
\mathcal{F}_a\left(y\right)=  \frac{\left( ay\right)^a}{\Gamma(a)} + \left( a y \right)^{\frac{a+1}{2}} I_{a-1}\left(2\sqrt{a y }\right)
\label{p(x) for gamma bursts and lambda equal 2 definition of mathcal F}
\end{eqnarray}
%
and $I_{\alpha}(z)$ is a modified Bessel function. Although (\ref{p(x) for gamma bursts and lambda equal 2}) has a rather complicated form, it has a simple expansion in terms of gamma distributions (\ref{gamma distribution definition a b}), namely 
\begin{eqnarray}
p_{\gamma}(x;2, \xi, a)= \frac{1}{e^a}\sum_{n=0}^{\infty} \frac{a^n}{n!}q_{\gamma}(x; n+a, \xi).
\label{p(x) for gamma bursts and lambda equal 2 gamma representation}
\end{eqnarray}
%


\section{Alternative way of obtaining the solution for exponential burst size distribution function analysed in Section \ref{Example exponential} and some of its properties  \label{Appendix F}}
The most convenient way to compute the inverse Laplace transform of $\hat{p}_{\epsilon}(s,t)$ (\ref{p hat in s space}) for $a = n \in \mathbb{N}$ is to rewrite $\hat{p}_{\epsilon}(s,t)=\hat{p}_{\epsilon, n}(s,t)$ as the following function of $\omega=e^{-\gamma t}$
\begin{eqnarray}
\hat{p}_{\epsilon, n}(s,t) &=& \hat{\tilde{p}}_{\epsilon, n}(s,\omega) = \left(\omega + \frac{1-\omega}{sb+1} \right)^{n} e^{-x_0 \omega s} \nonumber \\
&=& \sum_{i=0}^{n} {{n}\choose{i}}\frac{(1-\omega)^{i}\omega^{n-i}}{(sb+1)^{i}} e^{-x_0 \omega s}. 
\label{p hat in s space appendix}
\end{eqnarray}
According to Eq. (\ref{gamma distribution definition a b}), the inverse Laplace transform of $(sb+1)^{-i}$ is a gamma distribution with parameters $i$ and $b$, whereas the the inverse Laplace transform of unity is a Dirac delta function. Hence, for $x_0=0$, from (\ref{p hat in s space appendix}) we immediately obtain Eq. (\ref{solution for p no hat x}). If $x_0\geq 0$, instead of $\tilde{p}_{\epsilon, n}(x,\omega)$ (\ref{solution for p no hat x}) we obtain the more general solution $\tilde{p}_{\epsilon, n}(x,\omega; x_0)$. From the well-known properties of Laplace transform it follows that $\tilde{p}_{\epsilon, n}(x,\omega; x_0)$ has identical functional form as $\tilde{p}_{\epsilon, n}(x,\omega)$ (\ref{solution for p no hat x}), but the $x$ variable is replaced by $\xi=x-x_0 \omega$, i.e, $\tilde{p}_{\epsilon, n}(x,\omega; x_0) = \tilde{p}_{\epsilon, n}(x-x_0 \omega,\omega)$.

The explicit form of $\tilde{p}_{\epsilon, n}(x,\omega)$ (\ref{solution for p no hat x}) can be also obtained in an alternative way. From Eq. (\ref{ln pdf series expansion of Laplace transform epsilon}) for $a=n\in\mathbb{N}$ and $e^{-\gamma t}=\omega$ we have 
\begin{equation}
\frac{1}{\hat{p}(s \omega )} = \left( s \omega +\frac{1}{b} \right)^{n}  = \omega^{n} \sum_{k=0}^{n} {n \choose k} \frac{s^{n-k}}{b^k \omega^k}.
\label{1 over p in s space for exponential bursts}
\end{equation}
By applying the inverse Laplace transform to (\ref{1 over p in s space for exponential bursts}), we obtain
\begin{equation}
\frac{1}{\omega}  q\left(\frac{x}{\omega}\right) = \omega^{n} \sum_{k=0}^{n} {n \choose k} \frac{\delta^{(n-k)}(x)}{b^k \omega^k}.
\label{1 over p in x space for exponential bursts}
\end{equation}
For $x_0=0$, $\tilde{p}_{\epsilon, n}(x,\omega) $ is a convolution of $p(x)$ (\ref{gamma distribution definition a b}) and $  q(x/\omega)/\omega$ (\ref{1 over p in x space for exponential bursts}), cf. Eq. (\ref{t dep prob as convolution}). Invoking Eq. (\ref{convolution of delta derivatives}), we see that in order to obtain $\tilde{p}_{\epsilon, n}(x,\omega)$ we need to compute derivatives of $p(x)$ (\ref{gamma distribution definition a b}). However, one has to remember, that $p(x)$ is in fact a distribution, equal to $\Theta(x) q_{\gamma}(x; a, b)$, and not just an ordinary function $q_{\gamma}(x; a, b)$ (here by $\Theta(x)$ we denote a Heaviside step function). Therefore, in order to compute the $m$-th derivative of $p(x)$, Eq.  (\ref{distribution derivative}) should be invoked.

It is also convenient to rewrite Eq. (\ref{solution for p no hat x}) in a more compact form
\begin{eqnarray}
\tilde{p}_{\epsilon, n}(x,\omega) &=& \omega^n \delta(x) + \tilde{W}_{\epsilon, n}(x,\omega) e^{-\frac{x}{b}},
\label{solution for p no hat x appendix}
\end{eqnarray}
where 
\begin{eqnarray}
\tilde{W}_{\epsilon, n}(x,\omega) &=& \sum_{i=1}^{n} {{n}\choose{i}}\frac{(1-\omega)^{i}\omega^{n-i}}{(i-1)!b^{i}}x^{i-1}.
\label{solution for p no hat x definition of W}
\end{eqnarray}
Explicitly, for $n=0,1, 2$ and $3$, we have
\begin{eqnarray}
\tilde{W}_{\epsilon, 0}(x,\omega) &=& 0,
\label{n 0 special case solution}
\end{eqnarray}
\begin{eqnarray}
\tilde{W}_{\epsilon, 1}(x,\omega) &=& \frac{1-\omega}{b},
\label{n 1 special case solution}
\end{eqnarray}
\begin{eqnarray}
\tilde{W}_{\epsilon, 2}(x,\omega) &=& \frac{(1-\omega)^2}{b^2}x + 2\frac{(1-\omega) \omega}{b},
\label{n 2 special case solution}
\end{eqnarray}
\begin{eqnarray}
\tilde{W}_{\epsilon, 3}(x,\omega)&=&  \frac{(1-\omega)^3}{2 b^3}x^2  + 3\frac{(1-\omega)^2 \omega}{b^2}x \nonumber \\ &+&   3\frac{(1-\omega) \omega^2}{b}.
\label{n 3 special case solution}
\end{eqnarray}
Clearly, $\tilde{p}_{\epsilon, n}(x,\omega)$ as given by (\ref{solution for p no hat x}), (\ref{solution for p no hat x}) or by Eqs. (\ref{n 0 special case solution})-(\ref{n 3 special case solution}) has correct $\omega \to 1$ ($t \to 0$) and $\omega \to 0$ ($t \to \infty$) limits, as expected.

The validity of (\ref{solution for p no hat x}) for each $n$ can also be verified directly by inserting $\tilde{p}_{\epsilon, n}(x,\omega)$  to Eq. (\ref{unregulated t 
dependent ME of Friedman}). To deal with the term proportional to $\delta(x)$, one should invoke the identity $x \delta(x) \equiv 0$, in order to avoid differentiation of the distribution. Also, at $t=0$ ($\omega=1$), the initial condition (\ref{initial condition x}) is satisfied, and in the $t\to \infty$ ($\omega \to 0$) limit, only the $i=n$ term of (\ref{solution for p no hat x}) survives, leading again to gamma distribution (\ref{gamma distribution definition a b}). Also, it could be checked that $\tilde{p}_n(x,\omega)$ functions are correctly normalized for arbitrary $n$ and $\omega$. The weight of the  $\delta(x)$ term, equal to $\omega^n$, vanishes more rapidly for larger values of $n$.

It can be also checked that $(\tilde{p}_{\epsilon, 1}\ast \tilde{p}_{\epsilon, 1}) (x)= \tilde{p}_{\epsilon, 2}(x)$, $(\tilde{p}_{\epsilon, 1}\ast \tilde{p}_{\epsilon, 2})(x)= \tilde{p}_{\epsilon, 3}(x)$, etc., and that $\tilde{p}_{\epsilon, n}(x,\omega)$ (\ref{solution for p no hat x}) is a $n$-fold convolution of $ \tilde{p}_{\epsilon, 1}(x)$, i.e., it obeys Eq. (\ref{t dep prob in x as the n th convolution}).

The $k$-th moment of the probability distribution $\tilde{p}_n(x,\omega)$ (\ref{solution for p no hat x}) is equal to
\begin{eqnarray}
\tilde{\mu}^{(\epsilon, n)}_k(\omega) &\equiv & \int^{\infty}_{0} x^k \tilde{p}_{\epsilon, n}(x,\omega) dx \nonumber \\ &=&
\sum_{i=1}^{n} A^{(n)}_i b^{k+i}\Gamma(k+i),
\label{k th moment of solution for p no hat x}
\end{eqnarray}
%
%
where
\begin{equation}
A^{(n)}_i \equiv {{n}\choose{i}}\frac{(1-\omega)^{i}\omega^{n-i}}{(i-1)!b^{i}}
\end{equation}

In particular, we obtain
\begin{eqnarray}
\tilde{\mu}^{(\epsilon, n)}_1 &=& n b (1-\omega),\nonumber \\ 
\tilde{\mu}^{(\epsilon, n)}_2 - \left( \tilde{\mu}^{(\epsilon, n)}_1\right)^2 &=& n b^2 (1-\omega^2) \nonumber \\ &=&  \tilde{\mu}^{(\epsilon, n)}_1 b (1+\omega),
\end{eqnarray}
in agreement with Eq. (\ref{cumulant constant model parameters epsilon}). 
Note that the expression for $\tilde{\mu}^{(\epsilon, n)}_1$ obtained here is identical to that obtained in Ref. \cite{shahrezaei2008analytical} within the corresponding discrete model, whereas the expressions for the variance differ, although the difference is small if only  $b \gg 1$. \\


\bibliography{bibliography}

\end{document}